\documentclass[12pt,preprint]{aastex}
\usepackage{graphicx,color}
\usepackage{gensymb}
\shorttitle{BIAS AND THE NIRB POWER SPECTRUM}
\shortauthors{}
\begin{document}
\title{The Cosmic Near Infrared Background III: Fluctuations,
 Reionization and the Effects of Minimum Mass and Self-regulation}%
\author{%
   Elizabeth R. Fernandez$^1$,    Ilian T. Iliev$^2$ , Eiichiro Komatsu$^{3,4}$, 
\& Paul R. Shapiro$^3$
}%
\affil{%
$^1$Univ Paris-Sud, Institut d'Astrophysique Spatiale, UMR8617, 91405 Orsay Cedex, France \\
 $^2$Astronomy Centre, Department of Physics \& Astronomy, Pevensey II
 Building, University of Sussex, Falmer, Brighton BN1 9QH, United Kingdom\\
  $^3$Texas Cosmology Center and the Department of Astronomy, The University 
of Texas at Austin, 1 University Station, C1400, Austin, TX 78712\\
$^4$ Institute for the Physics and Mathematics of the Universe (IPMU), University of Tokyo, Chiba 277-8582, Japan
 }%
\email{%
  $^1$Elizabeth.Fernandez@ias.u-psud.fr
}%

\begin{abstract}
Current observations suggest that the universe was reionized sometime  before $z\sim 6$. One way to observe  this epoch of the universe is through the Near Infrared Background (NIRB),  which contains information about galaxies which may be too faint to be observed individually. We calculate the angular power spectrum ($C_l$) of the NIRB fluctuations caused by the distribution of these galaxies.  Assuming a complete subtraction of any post-reionization component, $C_l$ will be dominated by galaxies responsible for completing reionization (e.g., $z\sim 6$).  The shape of $C_l$ at high $l$ is sensitive to the  amount of {\it{non-linear}} bias of dark matter halos hosting  galaxies.  As the non-linear bias  depends  on the mass of these halos,  we can use the shape of $C_l$ to infer typical masses of  dark matter halos responsible for completing reionization. We extend our previous study by using a higher-resolution $N$-body simulation, which can resolve halos down to  $10^8~M_\sun$.  We also include improved radiative transfer, which allows for the  suppression of star formation in small-mass halos due to  photo-ionization heating.  As the non-linear bias enhances the dark-matter-halo power  spectrum on small scales, we find that $C_l$ is steeper for  the case with a complete suppression of small sources or  partial suppression of star formation in small halos (the minimum galaxy mass is $M_{\rm min}=10^9~M_\sun$ in ionized regions and  $M_{\rm min}=10^8~M_\sun$ in neutral regions) than the case in which these small halos were unsuppressed.  In all cases, we do not see a turn-over toward high $l$ in  the shape of $l^2 C_l$.
\end{abstract}  

\section{INTRODUCTION}
\label{sec:introduction}

Probing the beginnings of star and galaxy formation in the early universe is one of the 
goals of modern cosmology.  These high redshift stars provided a wealth of ionizing photons, 
which caused all of the hydrogen in the universe to be ionized, a process called reionization.
Therefore, understanding high-redshift star formation is closely coupled with our understanding of reionization history.  Recent
measurements of the polarization of
the Cosmic Microwave Background have shown that reionization started
early and was  
extended in time, with an equivalent instantaneous reionization redshift
of $z \sim 11$  \citep{kogut/etal:2003,spergel/etal:2003,spergel/etal:2007,
page/etal:2007,dunkley/etal:2008,komatsu/etal:2008}.  Therefore, we expect there
to be significant star formation before this.

With improving observations, there are many ways to be 
able to observe these first few generations of stars and galaxies.  We can now observe
high-redshift ($z>6$) star formation via galactic surveys, which provide
 statistical properties of high-redshift sources.  These surveys 
provide a wealth of information on early star formation; however, they only 
probe the ``tip of the iceberg,'' e.g., a small fraction of
the whole
population, mainly those galaxies 
that are bright enough to be identified in surveys. {\footnote{Those exceptionally bright galaxies also might be missed in surveys, because they are rare enough that they will not be found within the survey's limited area \citep{trenti/stiavelli:2008}}}.    

Another way to observe these high-redshift galaxies
is to look for the {\it diffuse} background light
originating from them. Ultra-violet photons produced by these early
galaxies at $z\gtrsim 6$
would be redshifted into the near infrared band ($\lambda\gtrsim
1~\mu$m). Therefore, any background above and beyond that from the
low-redshift galaxies could be attributable to these early
galaxies.
Observing  
the Near Infrared Background (NIRB) has the benefit of probing a population 
of galaxies {\it{as a whole}}, not just the unusually
bright 
objects which can be detected individually.
For this purpose, one can observe the mean NIRB \citep{santos/bromm/kamionkowski:2002,
magliocchetti/salvaterra/ferrara:2003,salvaterra/ferrara:2003,
cooray/yoshida:2004,madau/silk:2005, fernandez/komatsu:2006}  as well as 
fluctuations  \citep{kashlinsky/odenwald:2000,kashlinsky/etal:2002,
kashlinsky/etal:2004,kashlinsky/etal:2005,kash/etal:2007,kashlinsky/etal:2007,
kashlinskyb/etal:2007c,kashlinsky:2005,magliocchetti/salvaterra/ferrara:2003,
odenwald/etal:2003,cooray/etal:2004,matsumoto/etal:2005,thompson/etal:2007a,
thompson/etal:2007b, fernandez/etal:2010}. 

In our previous work \citep{fernandez/etal:2010}, we calculated
the angular power spectrum of the NIRB fluctuations using
an $N$-body simulation to trace the large-scale structure
and the formation of galactic halos, 
coupled with radiative transfer simulations of reionization.  
Combining this numerical data with the latest analytic modeling of
the internal, unresolved properties of radiation processes
such as stellar emission, Lyman-$\alpha$ line, two-photon emission,
free-free and free-bound emission from the formed galaxies
\citep{fernandez/komatsu:2006}, we analyzed the effects of  
changing the star formation efficiency $f_*$, the escape fraction
$f_{\rm esc}$\footnote{Here, by the ``escape fraction,'' we
refer to a fraction of hydrogen-ionizing photons escaping from the
galaxies into the intergalactic medium. We ignore dust extinction.}, 
and the mass and metallicity of the stars on the NIRB
angular power spectrum. 

Our previous simulation was able to resolve dark matter
halos down to 
the minimum mass of $M_{\rm min}=2.2\times 10^9~M_\sun$. In the 
current paper, we extend this analysis by improving the mass
resolution by more than an order of magnitude to $M_{\rm
min}=10^8~M_\sun$. We also include a new physical effect: the
suppression of star formation in small-mass halos due to 
photo-ionization heating (the so-called ``Jeans-mass filtering'',
which prevents baryons from collapsing into small dark matter
halos 
\citep[e.g.,][]{shapiro/etal:1994,2007MNRAS.376..534I}).

The structure of this paper is as follows. The simulations used for this 
analysis are described in \S~\ref{sec:simulation}. The fiducial stellar 
model used is discussed in \S~\ref{sec:stellarpop}, and our results are 
presented in \S~\ref{sec:results}. This model is then expanded upon in 
\S~\ref{sec:pops}.  We compare our results to observations in \S~\ref{sec:obs},
 and conclude in \S~\ref{sec:conclusions}.

\section{THE SIMULATION}
\label{sec:simulation}

In this paper, we use a subset of the simulations presented in \citet{iliev/etal(2011)}, discussed below. 
These new simulations have a significantly higher mass and spatial resolution 
than the ones we previously used in \citet{fernandez/etal:2010}. The minimum 
halo mass resolved (with 20 particles or more) is $10^8 \; M_{\sun}$ (vs. 
$\sim2 \times 10^9 \; M_{\sun}$ before). This much higher resolution allows 
us to explore the effect of the minimum halo mass of the halos that produce 
the ionizing photons on the angular power spectrum. Importantly, we can also
study the effects of the Jeans-mass filtering of low-mass halos due to the 
elevated temperature of the ionized intergalactic medium (IGM) compared to the 
neutral phase. This higher temperature suppresses the future formation of 
dwarf galaxies and stops the fresh gas infall onto already-formed ones. 
Hence, these new simulations do not simply provide an incremental improvement
on the resolution, but in fact include a whole new population of sources 
which are dynamically regulated by the reionization process \citep[see also]
[for further motivation and details on our model]{2007MNRAS.376..534I}. 

In this work we use two box sizes, $114~h^{-1}~\rm Mpc=163$~Mpc, which 
we use as our fiducial simulations, and $37~h^{-1}~\rm Mpc=53$~Mpc. The 
latter are computationally much cheaper and are used to test a wider 
range of source properties. The simulation parameters are summarized 
in Table \ref{tab:sims}. Both sets of simulations have the same minimum 
halo mass resolution, $M_{\rm halo, min}=10^8\,M_\odot$. We assume a constant 
mass-to-light ratio for the halos, defined by a parameter $f_\gamma$, 
which gives the number of ionizing photons produced per
stellar atom which manage to escape from the galaxy into the IGM
\citep{2006MNRAS.369.1625I}. The  
low-mass ($M<10^9M_\odot$), suppressible sources and the high-mass 
($M>10^9M_\odot$), unsuppressed ones typically have different efficiencies 
$f_{\gamma,small}$ and $f_{\gamma,large}$, respectively, reflecting the fact 
that the low-mass halos are expected to have larger populations of Pop~III, 
metal-free stars, which are much more efficient emitters of ionizing photons
than are Pop~II stars. Our two fiducial simulations consider the cases of 
high efficiencies ($f_{\gamma,large}=10$, $f_{\gamma,small}=150$, "{Partial\_Supp\_HighEff}'') and low ones ($f_{\gamma,large}=2$, $f_{\gamma,small}=10$, 
``{Partial\_Supp\_LowEff}''). 

In addition to our two fiducial cases, we also consider two more 
complementary cases to investigate the more extreme assumptions of
either no suppression of low-mass sources by reionization (case
``{No\_Supp}'') or a complete suppression, i.e., assuming that
halos with mass below $10^9M_\odot$ never formed any stars (case
``{Complete\_Supp}''). Cases with high efficiencies, {Partial\_Supp\_HighEff},
{No\_Supp}, and {Complete\_Supp}, reach overlap of ionized
regions (by design) at 
approximately the same redshifts, $z_{ov}=8.3$, 8.6 and 8.3, respectively,
where overlap is defined as the time when the mean ionized fraction 
by mass reaches $99\%$. The corresponding integrated electron 
scattering optical depths are $\tau_{\rm es}=0.080$, 0.078 and 0.071, 
respectively, all in agreement with the WMAP5 results 
\citep{komatsu/etal:2008}. In contrast, the
low-efficiency case,  
{Partial\_Supp\_LowEff}, corresponds to a more extended reionization 
history and reaches overlap only by $z=6.7$, with correspondingly 
lower integrated optical depth of $\tau=0.058$, slightly below the 
$2-\sigma$ limit from the WMAP 5-year results. 

\begin{table}[t]
\begin{center}
\begin{tabular}{|l|l|l|l|l|l|l|l|l|l|}

\hline
Simulation Name & Box Size &  Minimum  & Suppression & $f_{\gamma,large}$ 
& $ f_{\gamma,small}$ & $z_{\rm ov}$ & $\tau$  \\
& (Mpc$^3$) &  Halo Mass ($M_\sun$) & & & & &  \\
\hline
{Partial\_Supp\_HighEff} & 163  & $10^8$ & Yes  &  10  & 150 &8.3 &0.080\\
{Partial\_Supp\_LowEff}   & 163  & $10^8$ & Yes  & 2  & 10 & 6.7&0.058\\
{No\_Supp}   & 53 & $10^8$ & No  & 0.4  & 6 & 8.6 &0.078\\
{Complete\_Supp} & 53  & $10^9$ & Yes - complete & 12 & 0 & 8.3 &0.071 \\
\hline
\end{tabular}

\caption{Radiative transfer simulations used in this work.
}%
\label{tab:sims}
\end{center}
\end{table}

Throughout this paper, we use the cosmological parameters
($\Omega_{\rm m}$, $\Omega_\Lambda$, $\Omega_{\rm b}$, $h$)=(0.27, 0.73, 0.044, 
0.7) which are based on the WMAP 5-year results combined with the other
available constraints \citep{komatsu/etal:2008}.  

\section{Modeling the Stellar Population}
\label{sec:stellarpop}

We model the ionizing sources in the radiative transfer simulations based 
on the collapsed halos identified in the structure formation simulations, 
assuming a constant mass-to-light ratio.  This is defined by a variable 
$f_\gamma$, which quantifies how many ionizing photons escape the parent 
halo into the IGM over a timestep:
\begin{equation}
f_\gamma = f_* f_{\rm esc} N_i .
\end{equation}
Here, $N_i$ is the number of ionizing photons produced per
stellar atom, $f_{\rm esc}$  
is the escape fraction, and $f_*$ is the star formation efficiency, or the 
fraction of baryons within the halo that is converted into stars. 

For the 
purposes of modeling the Near Infrared Background, we also need to specify the 
spectral information as well as the spectral energy
distribution (SED) of the radiation, i.e., we need 
separate values for $N_i$ and $f_*$ above (still satisfying the overall 
$f_\gamma$ constraint). We shall take Population~II 
stars (with metallicity $Z=1/50 \; Z_\sun$) with a low escape fraction 
($f_{\rm esc} = 0.1$) obeying the following Salpeter initial mass spectrum
\citep{salpeter:1955} as our fiducial case:
\begin{equation}
f(m) \propto m^{-2.35}, 
\end{equation}
with mass limits of $m_1 = 3 \; M_\sun$ and $m_2 = 150 \; M_\sun$, for our 
entire range of redshifts.  The star formation efficiency ($f_*$) is then 
adjusted such that $f_\gamma$ is held fixed (i.e.,
the reionization constraint is satisfied).  The values of $f_*$ that are chosen for each value of $f_\gamma$ are
listed in Table \ref{tab:stellarpop}.
  
It is probably unphysical that 
only these stars will be present at all redshifts, since the very first
stars were metal-free, with quite different IMF and properties. Nevertheless,
we make this our fiducial model because the integrated NIRB angular power 
spectrum signal will likely be dominated by the later stages of reionization{\footnote{Here, we focus on galaxies present during the era of reionization.  To observe these galaxies, we assume that the low redshift component ($z\lesssim6$) can be completely removed.  In reality, this is a complicated observation to make.} 
by which point the metal enrichment should be fairly complete. This also represents the case with the highest amplitude for the angular 
power spectrum among several tested previously \citep[see Figures 7 \& 8 
of][]{fernandez/etal:2010}.  We explore other stellar populations in 
\S~\ref{sec:pops}.

\begin{table}[t]
\begin{center}
\begin{tabular}{|l|l|l|l|l|l|l|l|l|l|}

\hline
$f_{\gamma}$ &Stellar Population& $f_{*}$ & $ f_{\rm esc}$   \\
\hline
 0.4   & Pop II Salpeter &$ 1.5\times 10^{-3}$& 0.1\\
 0.4 & Pop III Larson & $1.6\times 10^{-5}$& 1\\
 2   & Pop II Salpeter & $7.7\times 10^{-3}$& 0.1\\
 2  & Pop III Larson & $8.0\times 10^{-5}$& 1\\
 6 & Pop II Salpeter & $2.3\times 10^{-2}$& 0.1\\
 6 & Pop III Larson & $2.4\times 10^{-4}$& 1\\
10  & Pop II Salpeter &$ 3.8 \times 10^{-2}$ & 0.1\\
 10   & Pop III Larson & $4.0 \times 10^{-4}$ & 1\\
12 & Pop II Salpeter &  $4.6\times 10^{-2}$& 0.1\\
 12 & Pop III Larson & $4.8\times 10^{-4}$& 1\\
  150  & Pop II Salpeter & $5.8 \times 10^{-1}$& 0.1 \\
  150   & Pop III Larson & $6.0 \times 10^{-3}$& 1\\
\hline
\end{tabular}

\caption{The properties of the stellar populations.  Each population is defined by an escape fraction, $f_{\rm esc}$, which is set to either high ($1$) or low ($0.1$).  When paired with a stellar mass function (Salpeter or Larson) and stellar metallicity (Pop II or Pop III), a value of the star formation efficiency, $f_*$ is solved for in order to be consistent with reionization.  
}%
\label{tab:stellarpop}
\end{center}
\end{table}

We can then compute the source luminosities in the infrared bands following the 
methodology we presented previously in \citet{fernandez/komatsu:2006} and 
\citet{fernandez/etal:2010}. The total emission has two components, the 
direct stellar emission; and any light re-processed by the nebula, including 
the Lyman$-\alpha$ line, free-free and free-bound emission, and two-photon 
emission. Any ionizing photons that escape the halos are re-processed into 
nebular emission within the IGM. Therefore, as $f_{\rm esc}$ increases, the 
amount of ionizing photons that remain within the halo to produce nebular 
emission falls, causing the overall luminosity of the halo to fall. (These 
ionizing photons will then produce nebular emission from the IGM, but because 
the IGM is so diffuse, the angular power spectrum of the IGM is anywhere from 
2 to 7 orders of magnitude lower than that of the halo 
\citep{fernandez/etal:2010} and thus its contribution to the overall
NIRB fluctuations can be safely disregarded.)

These analytical calculations are then combined with the power spectrum
of the mass density of dark matter halos
obtained from our simulations to calculate the 3-dimensional
power spectrum of the luminosity density, $P_L\left(k=\frac{l}{r(z)},z\right)$.
  This 3-D power spectrum is then integrated over redshift
  to obtain the angular power spectrum,
which is our key observable quantity \citep[see
Eq.~(37) of][]{fernandez/etal:2010}:
\begin{equation}
C_l =
\frac{c}{(4\pi)^2}
\int_{z_{\rm min}}^{z_{\rm max}} \frac{dz}{H(z)r^2(z)(1+z)^4}
P_L\left(k=\frac{l}{r(z)},z\right),
 \label{eq:cl}
\end{equation}
where $r(z)=c\int^z_0 dz/H(z)$ is the comoving distance. 
The integration range is taken from $z_{\rm min}=6$ to
$z_{\rm max}=30$. The luminosity power spectrum is proportional to the
mass density power spectrum of dark matter halos, $P_M(k)$, as
$P_L(k)=(L/M)^2P_M(k)$. The constant of proportionality is the
luminosity-to-halo mass ratio given by \citep[see
Eq.~(9) of][]{fernandez/etal:2010}:
\begin{equation}
 \frac{L(z)}{M}=f_*\frac{\Omega_b}{\Omega_m}
\left\{\bar{l}^*(z)+(1-f_{\rm
 esc})\left[\bar{l}^{ff}(z)+\bar{l}^{bf}(z)+\bar{l}^{2\gamma}(z)+\bar{l}^{\rm Ly\alpha}(z)\right]\right\}.
\end{equation}
Here, $\bar{l}^i$ is the band-averaged luminosity of a given radiation
process $i$ per stellar mass calculated from the formulae of
\citet{fernandez/komatsu:2006}, where $i=*$ (stellar), $ff$ (free-free), $fb$ (free-bound), $2\gamma$ (two-photon) and
Ly$\alpha$ (Lyman-$\alpha$ line). It is important to recall two facts:
\begin{itemize}
 \item [1.] The power spectrum amplitude is proportional to
       $(L/M)^2\propto f_*^2$.
 \item [2.] The higher the escape fraction $f_{\rm esc}$ is, the smaller
       the power spectrum becomes.
\end{itemize}
Therefore, for a given reionization history (i.e., a given value of
$f_\gamma=f_*f_{\rm esc}N_i$), the power spectrum amplitude is maximized when
$f_*$ is maximized while $f_{\rm esc}$ is minimized \citep{fernandez/etal:2010}.

\section{RESULTS}
\label{sec:results}

\subsection{Shape of the Angular Power Spectrum}

In Figure~\ref{fig:ClHalo37}, we show the angular power spectra for the 
following four cases: 1.) A minimum dark-matter-halo mass
of $10^8 \; M_\sun$ with no suppression of the star
formation due to photo-ionization heating, a minimum
halo mass of $10^8 \; M_\sun$ with 2.) high and 3.) with  
low efficiency with suppression turned on, and 4.) a minimum
halo mass of 
$10^9 \; M_\sun$ (in other words, complete suppression of low mass
sources). Results are presented for the J-band, which is assumed to be a
rectangular bandpass from 1.1 to 1.4 microns.  

\begin{figure}
\centering \noindent
\plotone{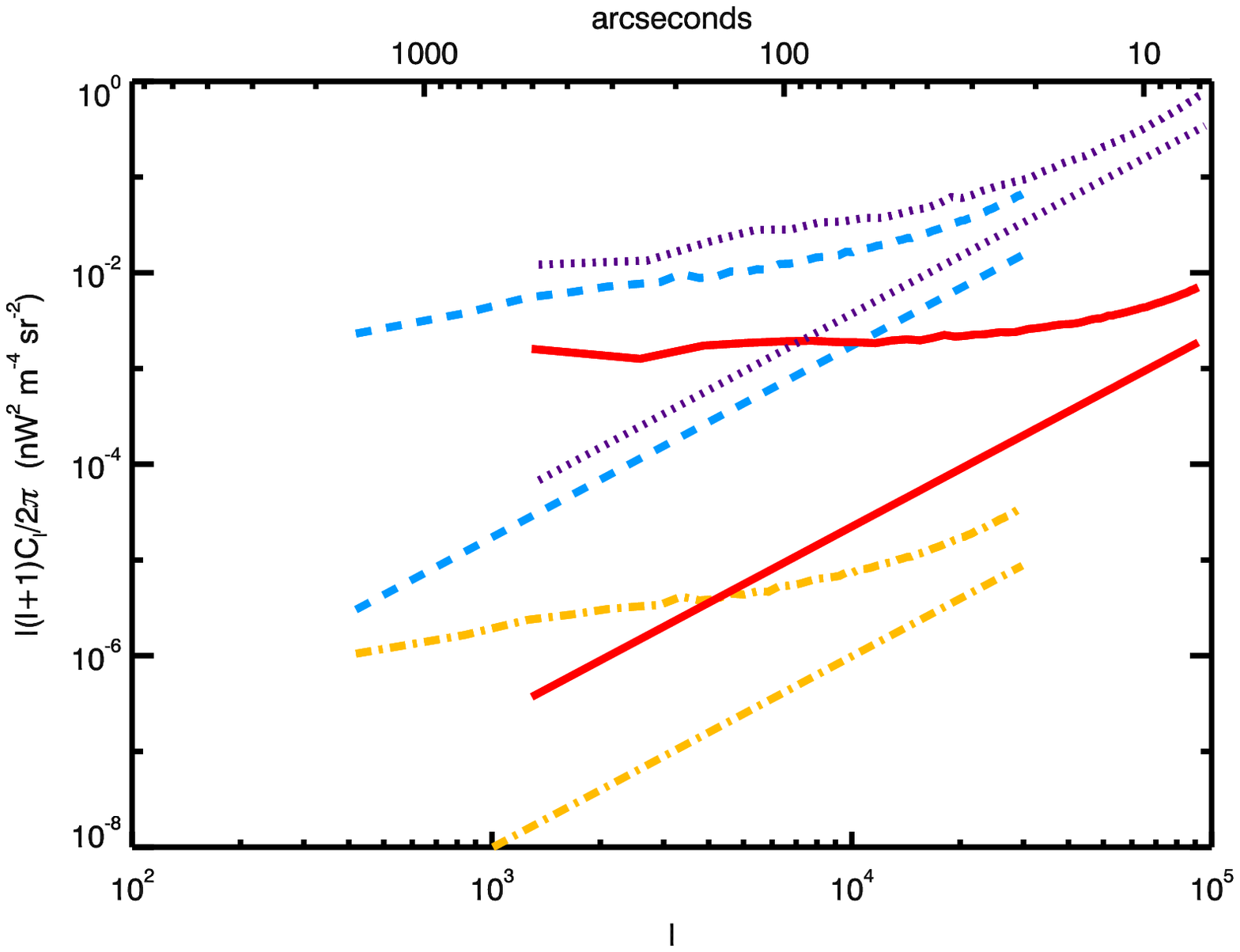}
\plotone{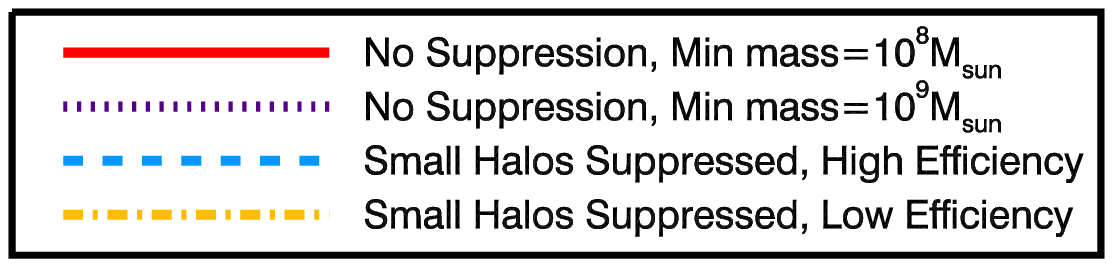}
\caption{
Angular power spectra of the NIRB fluctuation for Pop-II galaxies (with metallicity $Z=1/50 \; Z_\sun$) with a low escape fraction 
($f_{\rm esc} = 0.1$) and a Salpeter initial mass spectrum. The upper
 solid and dotted lines are for the minimum halo masses of $M_{\rm
 min}=10^8~M_\sun$ and $10^9~M_\sun$, respectively, with no
 suppression (the box size is 53~Mpc). The lower solid and dotted lines
 show the corresponding shot-noise contributions. The upper dashed and
 dot-dashed lines are for the high- and low-efficiency cases (see Table
 1), respectively, with suppression of  star formation in halos with
 $M<10^9~M_\sun$ in ionized regions (the box size is 163~Mpc).
The lower dashed and dot-dashed lines  show the corresponding shot-noise
 contributions.
The result is shown
 for  
the J-band (a rectangular bandpass from 1.1 to 1.4~$\mu$m).
  }
\label{fig:ClHalo37}
\end{figure}
\begin{figure}
\centering \noindent
\plotone{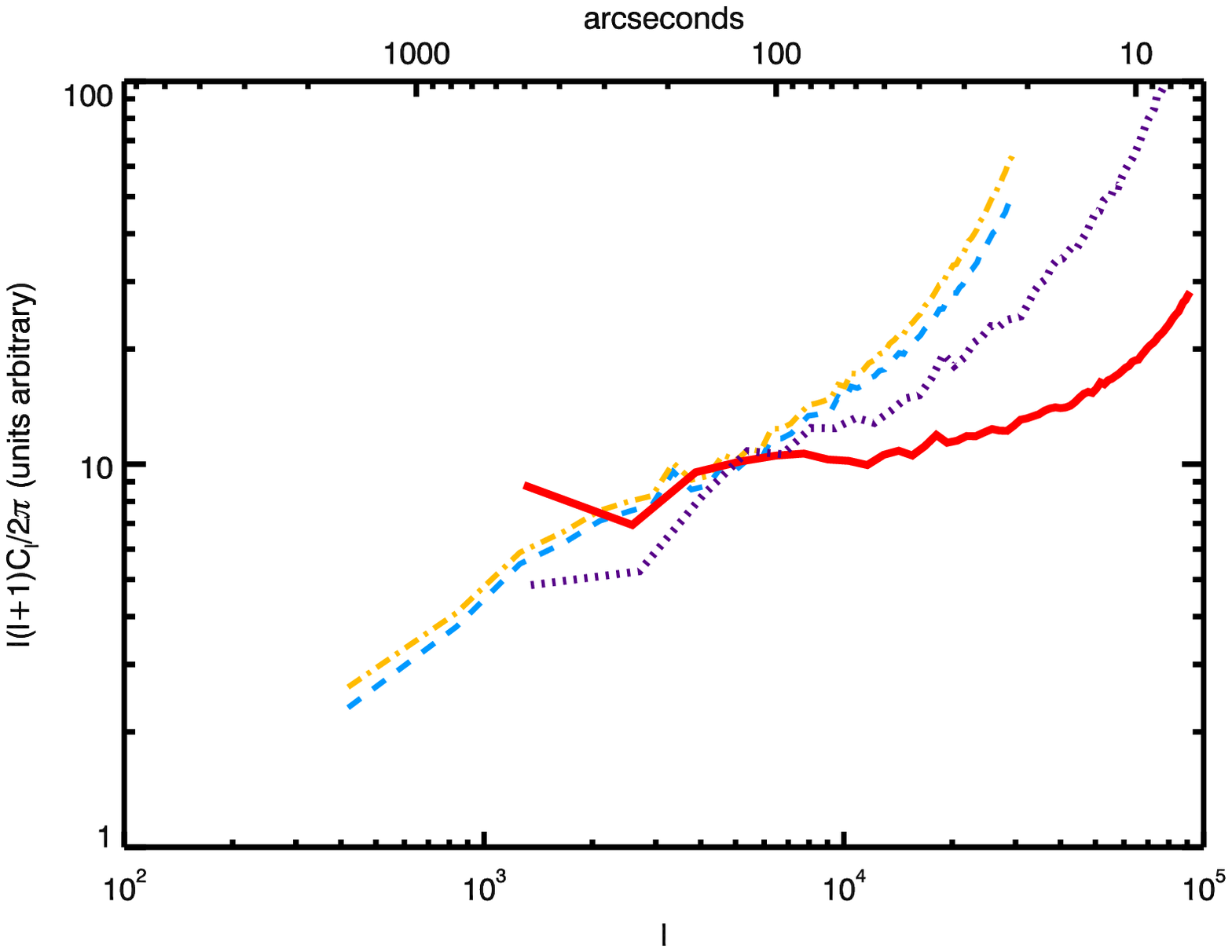}
\plotone{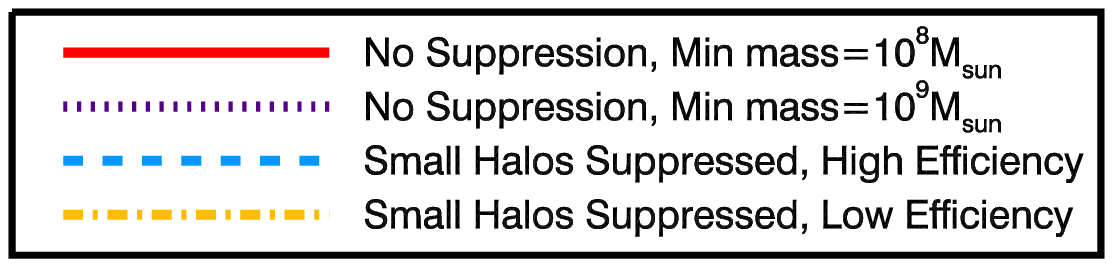}
\caption{Same as Fig.~1, but for the clustering component of the angular power spectrum ($C_l$ minus the shot-noise contribution), normalized to have the same amplitude at $l\approx 5000$. 
} 
\label{fig:yarbi}
\end{figure}

The previous calculations of the angular power spectrum 
\citep{kashlinsky/etal:2004,cooray/etal:2004} yielded
a turn-over of $l^2C_l$ at $l\sim 10^3$. 
Our calculation presented in
\citet{fernandez/etal:2010} did not confirm this. As we argued in that
paper, the previous calculations 
 did not take into account the non-linear halo
 bias, but used the linear bias, underestimating the power
 spectrum at high $l$. 
See the right panels of Figure~4 of
\citet{fernandez/etal:2010} for the $k$-dependence of the halo bias,
$b(k)$, or Figure~6 of \citet{iliev/etal(2011)}.  

The effect of non-linear bias will decrease as the minimum mass 
decreases, and this effect can be directly observable.
One outstanding question which was left unanswered in our
previous work is whether a lower minimum halo mass, $M_{\rm
min}=10^8~M_\sun$, can yield a turn over in $l^2C_l$. We  address
this question using the high-resolution simulation result presented in
this paper.
 In order to see more 
clearly the effect of the nonlinear halo bias
on the {\it shape} of the angular power spectrum, 
we show the clustered  
component (the angular power spectrum minus the shot-noise
contribution) in Figure~\ref{fig:yarbi}. 
To directly compare the shapes of these curves,
the clustered 
component of the angular power spectrum for each population
is normalized to be equal at $l \approx 5000$.  

As expected, we find that 
the angular power spectrum begins to flatten at high $l$ (around $\ell\sim10^3$) as the minimum
halo mass decreases.  This occurs simply because, as the 
minimum halo mass decreases, the non-linear halo bias decreases as well,
reducing the small-scale halo clustering power, and thus
causing the angular power spectrum to flatten.  
However, {\it{at no mass above $M_{\rm min}=10^8~M_\sun$ do we see a 
turnover of the angular power spectrum.}} 

The power spectrum is steepest for the (more physically realistic) cases 
with a partial suppression of small halos, and for the
(extreme) case with a complete suppression of $M<10^9~M_\sun$ halos.
Varying the source efficiencies (``{Partial\_Supp\_HighEff}''
versus ``{Partial\_Supp\_LowEff}'') does not  
change the shape of the power spectra noticeably (but does change its
amplitude). Therefore, {\it {the shape of the angular power spectrum gives 
information on the mass of the halos producing the NIRB fluctuations, with 
the steepest angular power spectrum resulting from populations either with 
a high minimum halo mass, or where smaller galaxies are
partially suppressed.}} 
Conversely, if 
low-mass, less biased sources are present and not suppressed by
reionization, the NIRB angular power spectrum would be much flatter.  

It is important to keep in mind that here, we assume a constant mass-to-light 
ratio for all halos, regardless of mass.  In reality, this is a simplified assumption, and it is probable
that smaller halos will have a smaller mass-to-light ratio.  One way to think of this is to relate the 
mass-to-light ratio of the halo
to the value of $f_{\gamma}$.  
If small halos have a smaller mass-to-light ratio than larger halos, $f_{\gamma,large}$ would fall as $f_{\gamma, small}$
rose, flattening the angular power spectrum.  We can see the effect of this type of trend if we compare our two cases without suppression (the angular power spectrum flattens if we move from $f_{\gamma,large}=12$ and $f_{\gamma,small}=0$ to $f_{\gamma,large}=0.4$ and $f_{\gamma,small}=6$).  This 
could be yet another variable that could affect the shape of the angular power spectrum.

\subsection{Amplitude of the Angular Power Spectrum}
  
Now, we can turn our attention to the amplitude of the angular power spectrum. 
The highest amplitude is produced if only the large sources are active 
(case ``{Complete\_Supp}''), i.e., the minimum source mass is $10^9 \; M_\sun$ 
and all smaller halos are completely suppressed. Why does this case with
a fewer, massive sources yield a higher NIRB angular power spectrum than
does a case when all halos are emitting (case ``{No\_Supp}'')?
The answer is simple: when only large  
halos are active, they are entirely responsible for reionization on their own. 
Consequently, 
in order to satisfy the reionization constraint,
a relatively high ionizing efficiency of $f_{\gamma,large}=12$ is 
required.
On the other hand, when all halos are emitting (i.e., when the minimum 
source mass is $10^8 \; M_\sun$), the massive source efficiency for the same 
overlap epoch is just $f_{\gamma,large}=0.4$. 
For a fixed initial mass spectrum (hence the number of
ionizing photons per stellar atom, $N_i$) and $f_{\rm esc}$, this
translates into 30 times smaller $f_*$, and thus 900
times smaller contribution to $C_l$. As a result, $C_l$ for this case is
dominated by small-mass halos.

Next, we consider the cases with partial
suppression of low-mass ($<10^9~M_\sun$) halos due to the
Jeans-mass filtering. 
In this instance, small halos will be suppressed in the vicinity of
large halos; thus, the story is similar to the case with a complete
suppression of  low-mass halos presented above. As some ionizing photons
come from low-mass halos, the efficiency for large-mass halos is
slightly smaller than the case above ($f_{\gamma,large}=10$ instead of
$12$; see Table~1).
This yields the NIRB
power spectrum amplitude that is similar to, but slightly lower than,
the case with $M_{\rm min}=10^9~M_\sun$ 
(``{Complete\_Supp}''), and significantly higher than the 
case with $M_{\rm min}=10^8~M_\sun$ without suppression
(``{No\_Supp}'').
Finally, a test-case scenario, in which source efficiencies
are so low that the overlap redshift,
$z_{ov}=6.7$, and the optical depth, $\tau=0.058$, are both low,
produces a significantly lower angular power spectrum than any of the
other cases.
 
\section{OTHER STELLAR POPULATIONS}
\label{sec:pops}

There are still significant uncertainties in the properties of the early stars 
and galaxies - their mass, metallicity, escape fraction, and star formation 
efficiency - and these can all be quite different from our fiducial case. The
dependence on  
these properties were discussed in depth in \citet{fernandez/etal:2010},
and will only be discussed briefly here.

There are multiple possibilities available for the stellar population. Along with 
our fiducial case, we will also discuss a case with Pop~III (zero-metallicity, 
$Z=0$) stars with a high escape fraction ($f_{\rm esc}=1$) and a top heavy, Larson 
initial mass spectrum \citep{larson:1998}:
\begin{equation}
f(m)\propto m^{-1}\left(1+\frac{m}{m_c}\right)^{-1.35},
\end{equation}
with $m_1 = 3M_\sun$, $m_2=500M_\sun$, and $m_c = 250M_\sun$. Our earlier results 
in \citet{fernandez/etal:2010} suggested that, while our fiducial case above 
(Pop~II stars with a Salpeter initial mass spectrum
and a low escape fraction) represented the 
limit with the highest amplitude of the NIRB angular power spectrum,
this new case  
represents the opposite limit of a low NIRB amplitude. Different combinations of escape 
fraction and star formation efficiency, paired with these mass functions and 
metallicities, will most likely yield amplitudes between these two limits.

We can further vary which populations are found in which type of source - high-mass 
and low-mass halos do not necessarily contain the same stellar populations.  The 
results are shown in Figure~\ref{fig:ClHalo37all}. In all cases, the steepest 
angular power spectrum results from those cases with the largest
minimum halo mass of $10^9 \; M_\sun$ as well as those with a partial
suppression, and  
the shallowest from the cases with the lowest minimum halo mass of $10^8 \; M_\sun$, 
although the detailed 
shapes and amplitudes vary significantly between different cases.

The highest NIRB angular power spectrum amplitudes results when the
larger sources contain the bright Population 
II stars with low escape fraction. For most  
cases, the brightest NIRB emission is produced when only the high-mass
sources are emitting.  In addition, the
 large-mass halos only case ``{Complete\_Supp}'' is roughly
equivalent to the case ``Partial\_Supp\_HighEff,'' in which
the small halos are suppressed and $f_{\gamma,large}$ is high.
As before, this can be explained by the value of
$f_{\gamma,large}$: angular power spectra with a high amplitude have a
high value of $f_{\gamma,large}$. Small halos do not contribute if 
they are dynamically suppressed.

Comparison between the top-left (Pop II Salpeter for all halos)
and top-right panels (Pop II Salpeter for large-mass halos and Pop III Larson
for small-mass halos) of Figure~3 would show this clearly. 
As the Pop-III-Larson case has $f_{\rm esc}=1$, and its $f_*$ is $\sim
10^2$ times smaller than that of the Pop-II-Salpeter case (see Table~2),
the small-mass halos are significantly dimmer in the top-right
panel. The dotted lines are identical in both panels, as these show
the case with $M_{\rm min}=10^9~M_\sun$ and thus the same Pop II Salpeter.
The suppressed cases, shown by dashed and dot-dashed lines in both panels, are also very similar. This is
because small-mass halos are suppressed anyway, and thus the results do not
change much by making small-mass halos dimmer.
On the other hand, the cases with no suppression (the solid lines) are very different: in the top-left
panel, the small-mass halos dominate the power spectrum and thus the
power spectrum is much flatter than the dotted line. However, in the
top-right panel, the small-mass halos are much dimmer and thus the power
spectrum is dominated by the large-mass halos again, making the shape of
the power spectrum as steep as the dotted line. (The amplitude is 
$900$ times smaller than the dotted line, as $f_{*,large}$ for the
solid line is 30 times smaller than that for the dotted line.) 

The one exception is when the massive sources are dimmer,
containing Pop III stars with a large escape fraction, while 
low-mass halos contain Pop II stars with a low escape fraction. In 
this case, adding brighter small halos is more than enough to compensate 
for having ``less efficient'' large halos, even when these small halos
are suppressed. However, because of the hierarchical  
nature of the cosmological structure formation, it is physically unlikely to 
have metal-enriched low-mass halos while having unenriched high-mass halos, 
and thus this case is the least plausible.

\begin{figure}
\centering \noindent
\plottwo{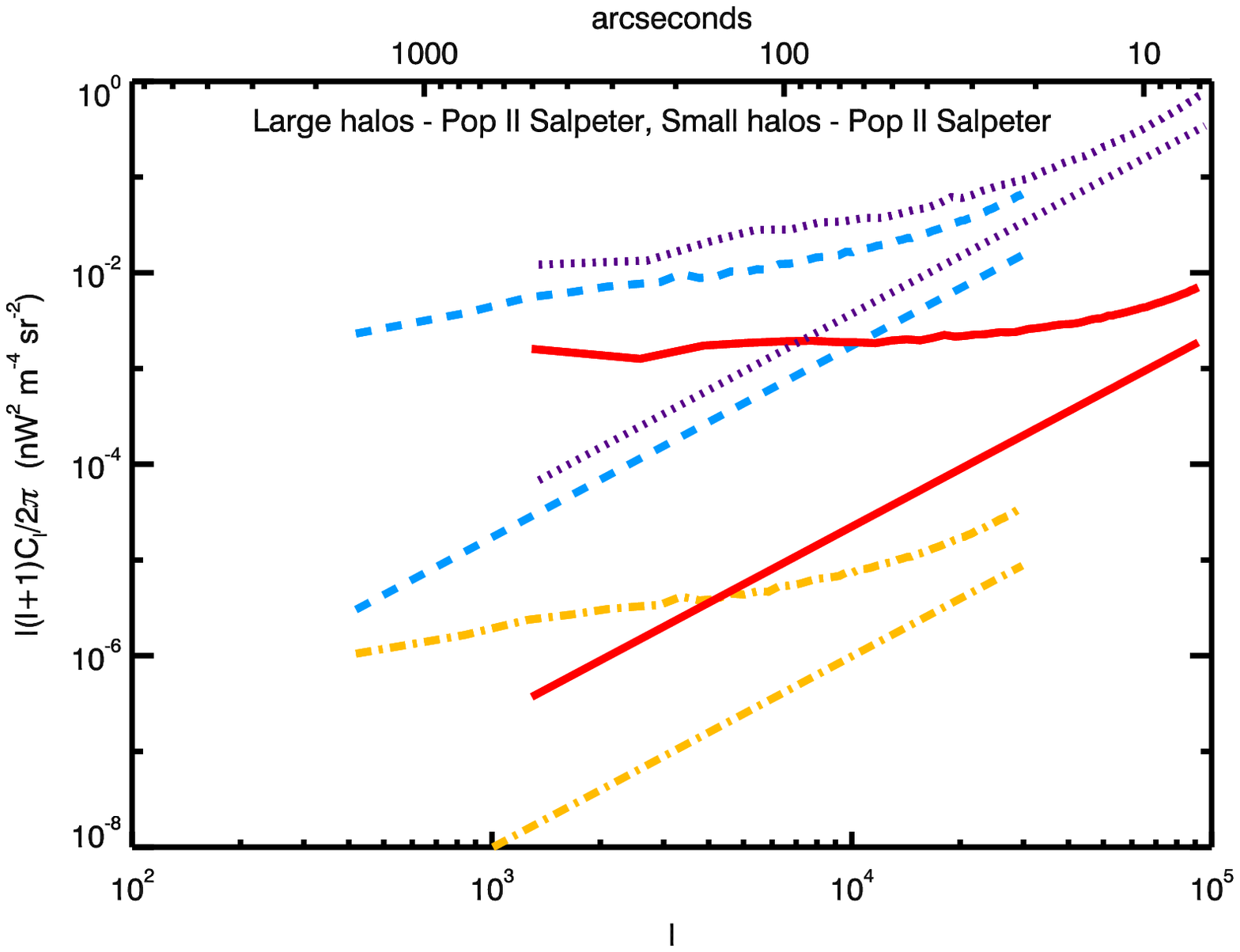}{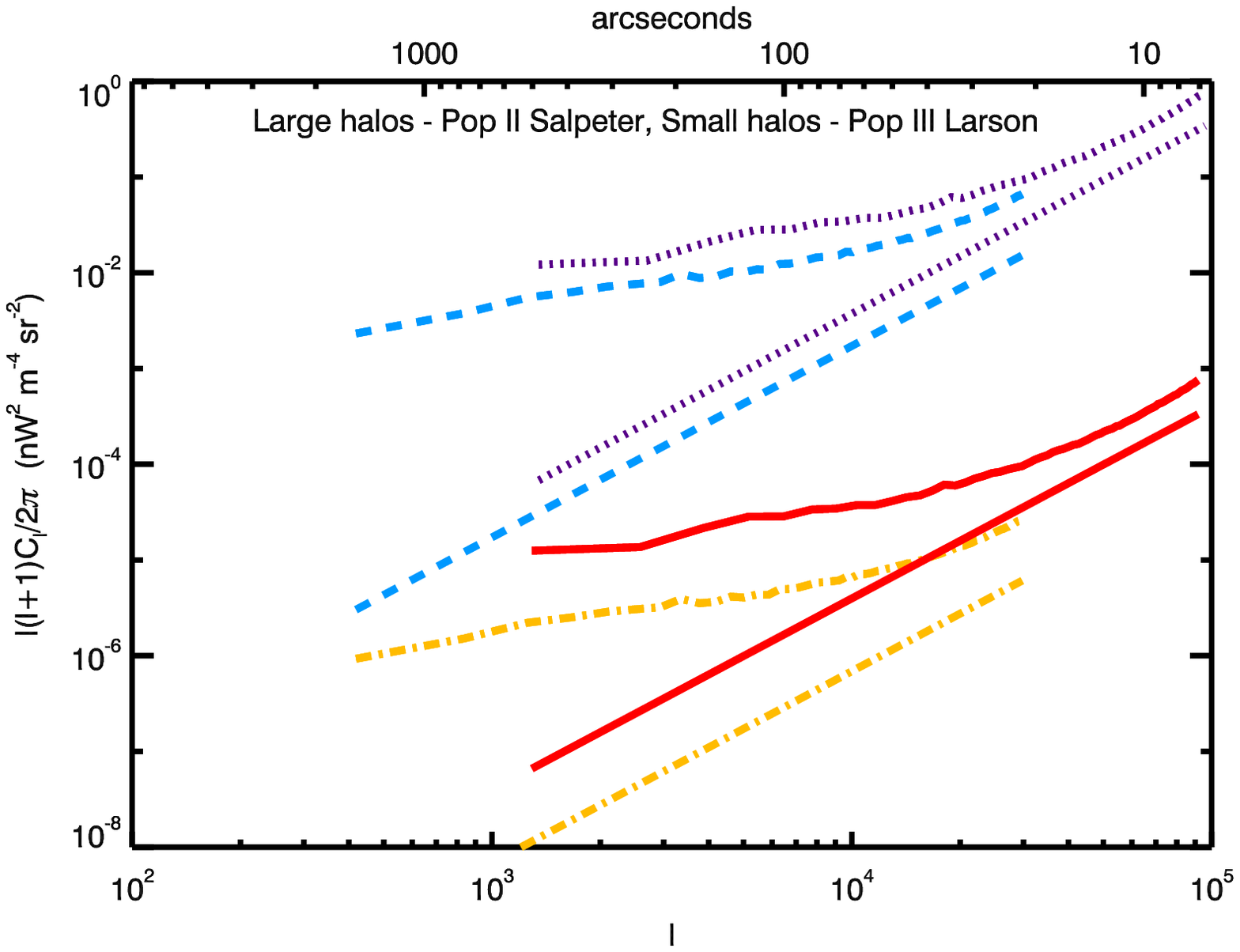}
\plottwo{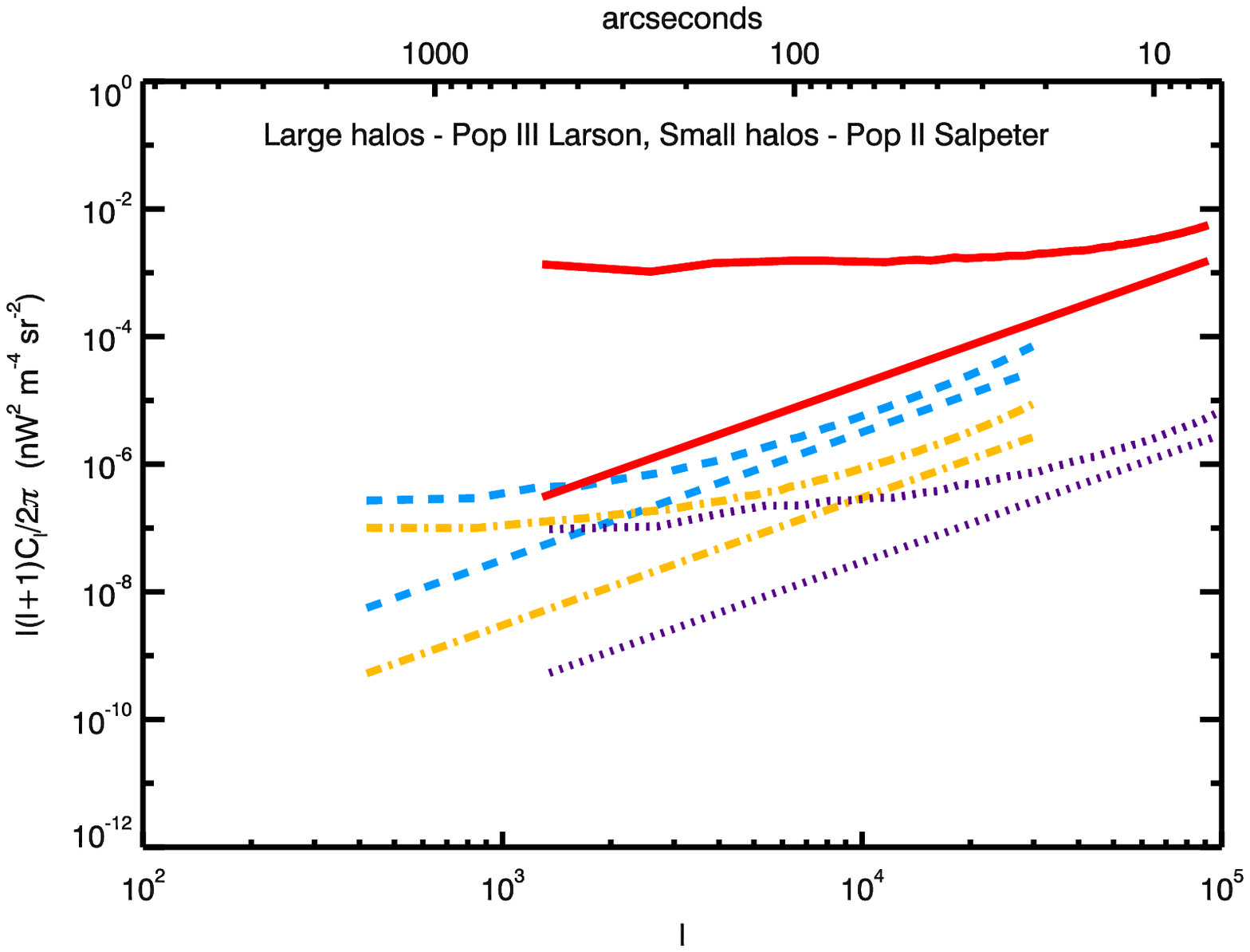}{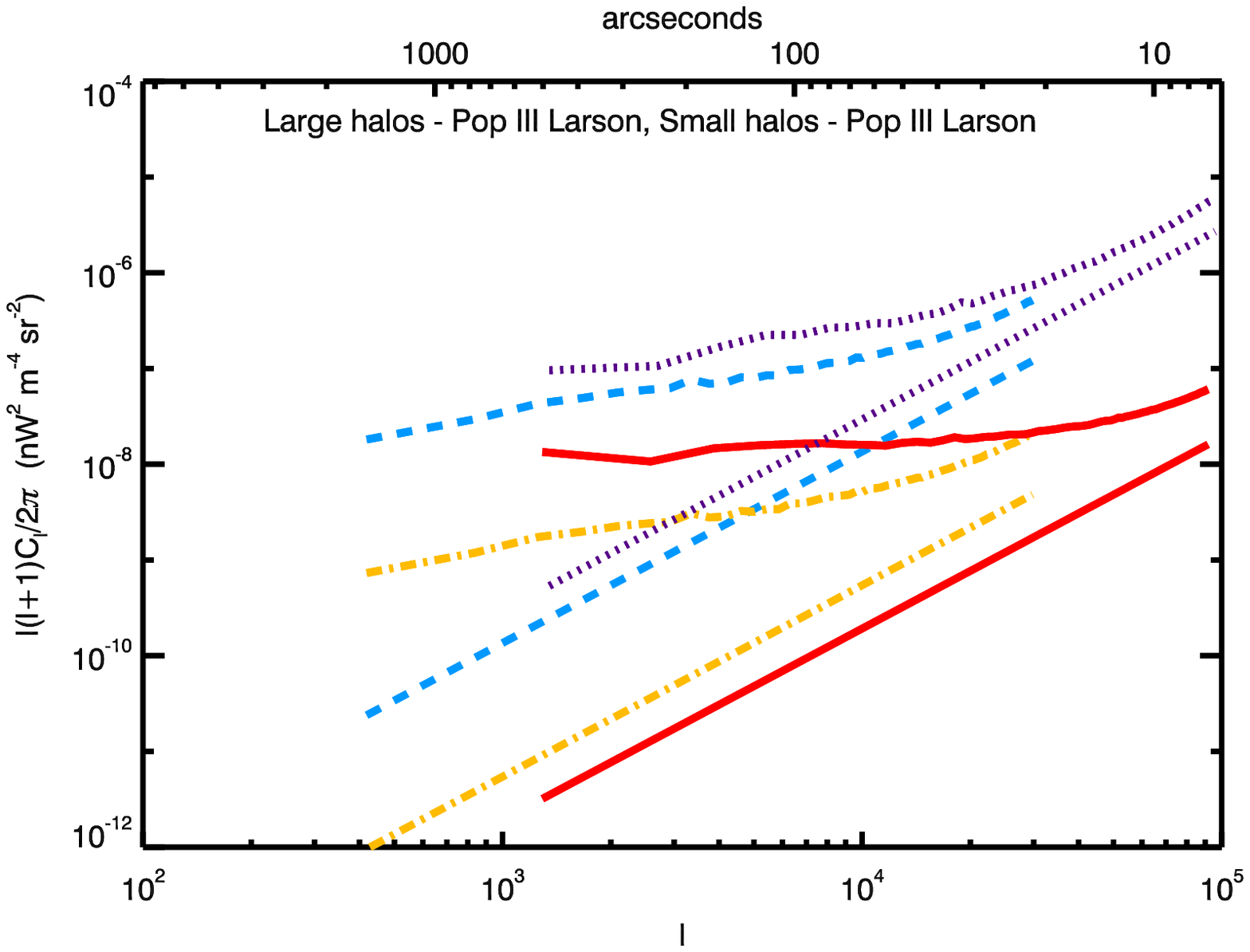}
\plotone{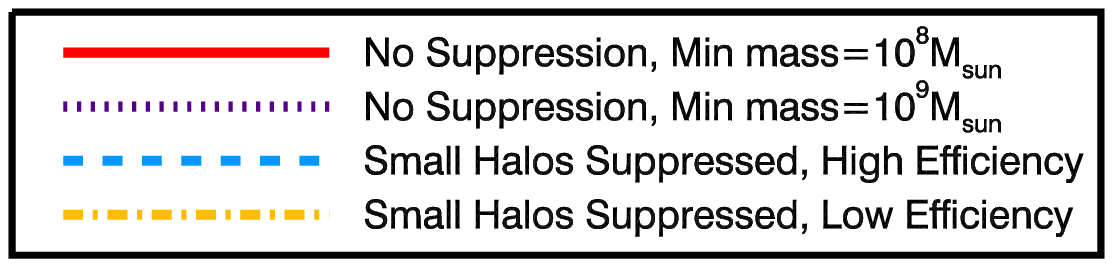}
\caption{Same as Fig.~1, but for various stellar populations.
Top left panel: the same population as Fig.~1 (Pop II Salpeter for all
 halos). Top 
 right panel: Pop II Salpeter for large-mass ($>10^9~M_\sun$) halos; Pop
 III Larson for small-mass ($<10^9~M_\sun$) halos. Bottom left panel:
 Pop III Larson for large-mass halos; Pop II Salpeter for small-mass
 halos. Bottom right panel: Pop III Larson for all halos.}
\label{fig:ClHalo37all}
\end{figure}

\section{OUR MODEL AND OBSERVATIONS}
\label{sec:obs}

There have been multiple recent attempts to measure the fluctuation power 
of the NIRB due to high-redshift, unresolved galaxies. 
Measurements of the diffuse background light
are always challenging, as they depend upon careful subtraction of all
foreground sources, which is 
complicated. For example, zodiacal light is a main contaminant, 
and no complete model for it is available. When dealing with fluctuations, 
however, the main uncertainties are due to low redshift ($z<6$) galaxies.  
Typically, a map with a certain limiting magnitude is used to mask out low 
redshift galaxies, and another component is used to account for galaxies that 
are too dim to be seen, assuming a certain evolution scenario. In order to 
obtain accurate results, one must be careful that these foreground components 
are successfully removed while still leaving enough of the map to perform 
accurate fluctuation analysis.

We have chosen to compare our theoretical predictions with
the results of several groups who presented 
measurements at a range of frequencies in the near infrared. The first is 
\citet{thompson/etal:2007a}, who used the Near Infrared Camera and Multi-Object 
Spectrometer (NICMOS) on the {\it{Hubble Space Telescope}} to measure 
fluctuations in the NIRB at 1.1 and 1.6 microns. They removed zodiacal light 
by dithering the camera, and removed foreground galaxies down to the ACS and 
NICMOS detection limits. \citet{matumoto/etal:2010} measured the fluctuations 
of the NIRB using the AKARI satellite at 2.4, 3.2, and 4.1 microns. The camera 
was rotated over several epochs to remove zodiacal light fluctuations. To remove 
foreground galaxies, they followed a procedure which included masking all pixels 
over $2-\sigma$, repeating the process ten times, and removing foreground sources 
using DAOFIND.  Finally, two groups analyzed data from the Infrared Array Camera 
(IRAC) on the {\it{Spitzer Space Telescope}} at 3.6, 4.5, 5.8 and 8 microns 
\citep{kashlinsky/etal:2005, kashlinsky/etal:2007, cooray/etal:2007}. 
\citet{kashlinsky/etal:2005} and \citet{kashlinsky/etal:2007} removed zodiacal 
light by taking observations 6 months apart in fields rotated by 180$\degree$.  
Both groups removed pixels $\gtrsim 4\sigma$ above the mean. To remove fainter 
sources, \citet{kashlinsky/etal:2005, kashlinsky/etal:2007} removed sources 
found by SExtractor convolved with the point spread function of IRAC.  
\citet{cooray/etal:2007}, on the other hand, masked the image to various 
magnitude limits and used known galaxies in the ACS catalog.

\begin{figure}
\centering \noindent
\plotone{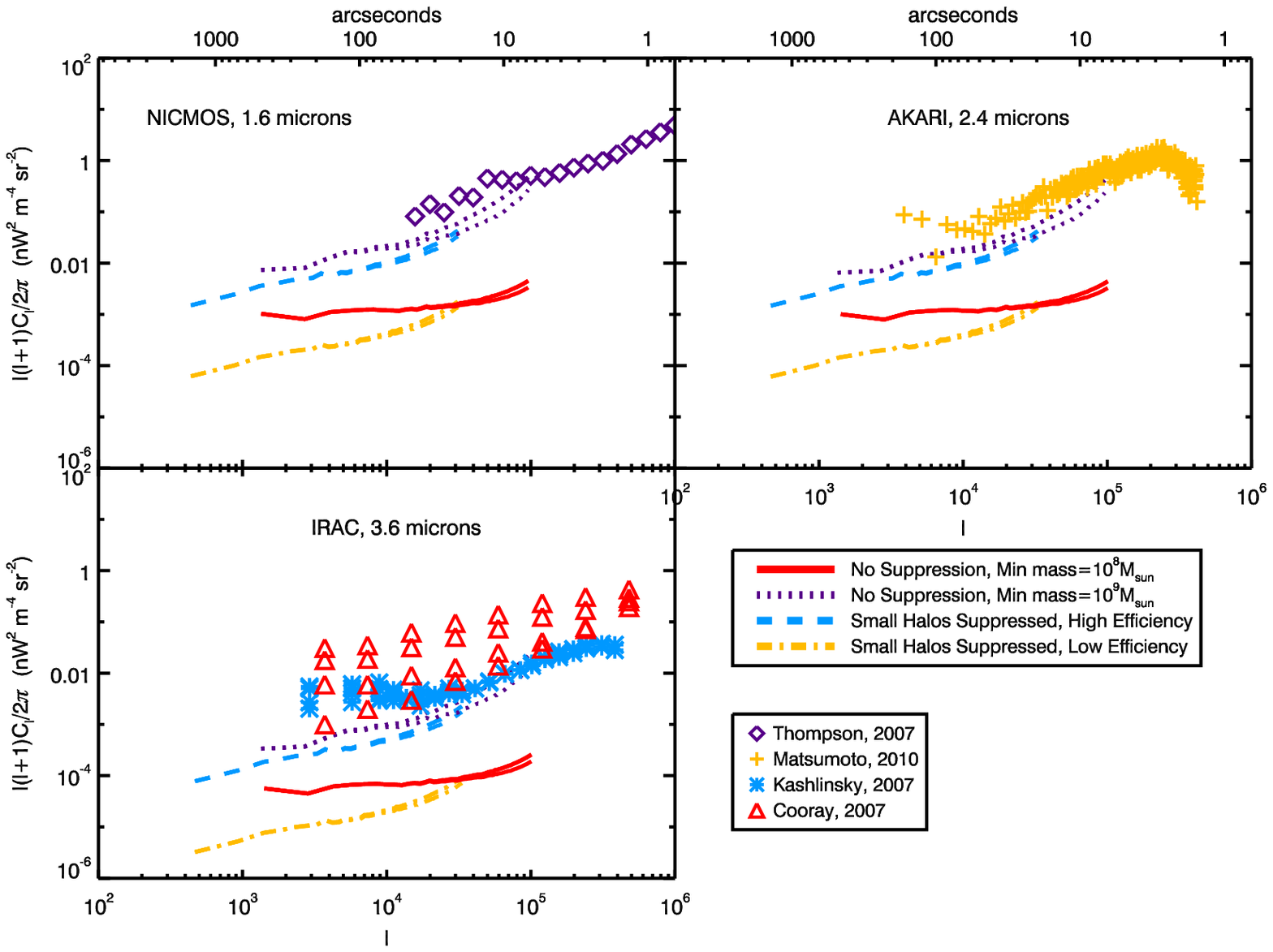}
\caption{Our fiducial model (Pop~II stars with a Salpeter initial mass
 spectrum and $f_{\rm esc}=0.1$) predictions 
for the NIRB angular power spectra  in various bands, compared to
the observational results at 1.6 microns from the NICMOS  
camera \citep[top-left panel]{thompson/etal:2007a}, 2.4 microns from AKARI 
\citep[top-right panel]{matumoto/etal:2010}, and 3.6 microns from IRAC 
\citep[bottom-left
 panel]{kashlinsky/etal:2005,kashlinskyb/etal:2007c,cooray/etal:2007}.   For our models,
 the top line in each case is without the shot-noise subtracted, and the bottom line is with our prediction
 for the shot-noise subtracted.  Note that in some cases (ie, the measurements for \citet{cooray/etal:2007}), 
 a higher model of shot-noise than ours is subtracted from the observations (namely, where the shot-noise is fit to the angular power spectrum at high multipoles, and the data may contain contributions from both high-$z$ and low-$z$ galaxies).}  
\label{fig:obs}

\end{figure}

In Figure~\ref{fig:obs}, we compare our fiducial
calculations (with Pop~II stars; a 
Salpeter initial mass spectrum; and $f_{\rm esc}=0.1$) to the
measured 
angular power spectra at 1.6
microns from the NICMOS camera  
\citep{thompson/etal:2007a}, 2.4 microns from AKARI \citep{matumoto/etal:2010}, 
and 3.6 microns from IRAC \citep{kashlinsky/etal:2005, kashlinsky/etal:2007,
cooray/etal:2007}. 
In all cases, the measurements are above our
calculations.
These measurements should probably be taken as upper limits
on the contributions from galaxies in $z\gtrsim 6$,
as there may be some lower redshift components or other foreground
contamination in these measurements. 
We also note that our fiducial model presented here has the
largest predicted amplitude of $C_l$; thus, if we were to change  
the population within our range of models satisfying the
reionization constraint, the predicted
amplitude of the angular power  
spectrum would decrease. 
In other words, 
these upper limits do not yet rule out any 
of our models. We also note that current observational limits are many orders 
of magnitude above that we would expect from our lowest magnitude of the 
angular power spectrum.  Many of our models are far beneath the sensitivity of current observations.  The upcoming experiment CIBER will push the sensitivity limit further \citep{cooray/etal:2009}, and may begin to be able to discern between various high-redshift populations.  Still, it is possible that many of our models will need extraordinarily sensitive instruments to be detected  (see figure 18 of \citet{fernandez/etal:2010}).

\section{CONCLUSIONS}
\label{sec:conclusions}

The NIRB mean and fluctuation signals can be a very powerful probe of
the high-redshift star formation ($z\gtrsim 6$). 
Building upon our previous work on this subject
\citep{fernandez/komatsu:2006,fernandez/etal:2010}, we have extended the
calculation of the angular power spectrum of the NIRB fluctuation by
improving the mass resolution of $N$-body simulations by an order of
magnitude. With the simulation now resolving the dark matter halo mass
down to $M=10^8~M_\sun$, we have confirmed our previous finding: due to
the non-linear halo biasing, the 
shape of $l^2C_l$ of the NIRB fluctuation does not exhibit a turn over. However, we do observe a flattening
of the shape as we lower the minimum halo mass (because the non-linear
bias decreases as masses go down), and thus the shape of
$l^2C_l$ can be used to infer typical masses of dark matter halos
hosting galaxies in a high redshift universe.

We have gone beyond simply increasing the mass resolution of the
simulation. For the first time, we consider cases with a 
radiative feedback suppressing the star formation in
low-mass galaxies with $M<10^9~M_\sun$, due to the Jeans-mass filtering in
the ionized and heated IGM. 
We find that, when low-mass sources are partially suppressed by 
photo-ionization heating, 
the predicted angular power spectrum becomes
similar to the one with a complete suppression of halos below $M_{\rm
min}=10^9~M_\sun$, yielding a steep power spectrum at high $l$. 
Therefore, the shape 
of the angular power spectrum at high $l$ can directly provide information on 
the typical mass of sources responsible for completing
reionization. 

The amplitude of the angular power spectrum is less robust,
as it depends on a number of parameters such as the stellar initial mass spectrum,
metallicity of stars, star formation efficiency, and escape
fraction. However, one robust feature is that it is largely determined
by the properties of the halo 
populations at late stages of reionization.  Therefore, the angular
power spectrum is higher if the stars produce more ionizing photons that
do not escape from the halo.  (Therefore, maximizing $f_*$ and $N_i$
while minimizing $f_{\rm esc}$ within the limitations of reionization.) 

Finally, we find that our predictions, all of which are tuned to
satisfy the reionization constraints, are below the current
measurements. Given that these measurements should probably be taken as
upper limits on the contributions from galaxies in $z\gtrsim 6$, we
conclude that our calculations are consistent with the current measurements.

\section*{Acknowledgments} 
We would like to thank Garrelt Mellema for helpful 
discussions. ERF was supported by the ANR program ANR-09-BLAN-0224-02.  ITI was supported by The Southeast Physics Network (SEPNet) and 
the Science and Technology Facilities Council grants ST/F002858/1 and
ST/I000976/1. 
PRS was supported by NSF grants AST-0708176 and AST-1009799, and NASA
grants NNX07AH09G, NNG04G177G, and NNX11AE09G. 
We acknowledge support from the National Science Foundation through TeraGrid 
resources provided by the Texas Advanced Computing Center (TACC) under grant 
number TG-AST090005. We also thank the TACC at The University of Texas
at Austin  
for providing HPC resources that have contributed to the research results reported 
within this paper (URL: http://www.tacc.utexas.edu).

\end{document}